\documentclass[draft]{agujournal2019}
\usepackage{url} %this package should fix any errors with URLs in refs.
\usepackage{bm}
\usepackage{lineno}
\usepackage[finalnew]{trackchanges}
\usepackage{soul}
\usepackage{amsmath}
\journalname{JGR: Space Physics}

\begin{document}

\title{Magnetic Field Draping in Induced Magnetospheres: Evidence from the MAVEN Mission to Mars}

%% ------------------------------------------------------------------------ %%
%
%  AUTHORS AND AFFILIATIONS
%
%% ------------------------------------------------------------------------ %%

\authors{A. R. Azari\affil{1}, E. Abrahams\affil{2}, F. Sapienza\affil{2}, D. L. Mitchell\affil{1}, J. Biersteker\affil{3}, S. Xu\affil{1}, C. Bowers\affil{4}, F. Pérez\affil{2}, G. A. DiBraccio\affil{5}, Y. Dong\affil{6}, \add[ver3]{S. Curry}\affil{1}} 

%authorship arrangements being finalized

\affiliation{1}{Space Sciences Laboratory, University of California, Berkeley, California}
\affiliation{2}{University of California, Berkeley, California}
\affiliation{3}{Massachusetts Institute of Technology, Cambridge, Massachusetts}
\affiliation{4}{University of Michigan, Ann Arbor, Michigan}
\affiliation{5}{NASA Goddard Space Flight Center, Greenbelt, Maryland}

\affiliation{6}{Laboratory for Atmospheric and Space Physics, University of Colorado, Boulder, Colorado}

\correspondingauthor{A. R. Azari}{azari@berkeley.edu}%{azari <at> berkeley <dot> edu}

\begin{keypoints}

\item We quantify Mars’ induced magnetic field response as compared to responses from the intrinsic crustal and direct solar wind magnetic fields.
\item We expand induced magnetosphere theory after showing twisted dayside fields depend on an underestimated induced effect and crustal fields. 
\item Twisted dayside draping is a precursor to, and aligns with, observations of the twisted magnetotail at Mars.

\end{keypoints}
%% ------------------------------------------------------------------------ %%
%
%  ABSTRACT and PLAIN LANGUAGE SUMMARY
%
%% ------------------------------------------------------------------------ %%

%% \begin{abstract} starts the second page

\begin{abstract}
The Mars Atmosphere and Volatile EvolutioN (MAVEN) mission has been orbiting Mars since 2014 and now has \add[ver2]{over 10,000}\remove[ver2]{tens of thousands of} orbits which we use to characterize Mars' dynamic space environment.
Through global field line tracing with MAVEN magnetic field data we find an altitude dependent draping morphology that differs from expectations of induced magnetospheres in the vertical ($\hat Z$ Mars Sun-state, MSO) direction.
We quantify this difference from the classical picture of induced magnetospheres with a Bayesian multiple linear regression model to predict the draped field as a function of the upstream interplanetary magnetic field (IMF), remanent crustal fields, and a previously underestimated induced effect. 
From our model we conclude that unexpected twists in high altitude dayside draping ($>$800 km) are a result of the \add[ver2]{IMF component in the $\pm \hat X$ MSO direction}\remove[ver2]{IMF angle in the X, Y MSO plane}.
We propose that this is a natural outcome of current theories of induced magnetospheres but has been underestimated due to approximations of the IMF as solely $\pm \hat Y$ directed. 
We additionally estimate that distortions in low altitude ($<$800 km) dayside draping along $\hat Z$ are directly related to remanent crustal fields.
We show dayside draping \add[ver2]{traces}\remove[ver2]{propagates} down tail and previously reported inner magnetotail twists are likely caused by the crustal field of Mars, while the outer tail morphology is governed by an induced response to the IMF direction.
We conclude with an updated understanding of induced magnetospheres which details dayside draping for multiple directions of the incoming IMF and discuss the repercussions of this draping for magnetotail morphology.

\end{abstract} 

%----------------------------------------------------------------------------------
\section*{Plain Language Summary}

Mars presents a dynamic and complex obstacle to the solar wind, a supersonic flow of magnetized plasma from the Sun.
This complexity is due to Mars' ionized upper atmosphere, or ionosphere, and a patchwork of strong crustal magnetic fields that rotate with the planet.
These factors perturb how the solar wind magnetic field drapes around Mars. 
The solar wind magnetic field is most often approximated as either eastward or westward directed as referenced by Mars' longitude.
For planets without internally generated magnetic fields like Venus it is expected that the solar wind magnetic field would drape symmetrically around the ionosphere.
However, spacecraft measurements have revealed a more complicated draping pattern at Mars including northward and southward directed draping. 
We show that these alterations can be explained by combination of crustal magnetic fields at low altitudes and by a previously underestimated effect of induced magnetospheres at high altitudes.  
These dayside draping alterations propagate globally throughout the Mars system, potentially explaining previously observed alterations in expected magnetotail observations. 
We conclude with an updated guide for how solar wind magnetic field draping occurs around induced obstacles and suggest that this is a unifying phenomenon of planetary and solar wind interactions.
%
%
% The Plain Language Summary should be written for a broad audience,
% including journalists and the science-interested public, that will not have a background in your field.

%----------------------------------------------------------------------------------
\section{Introduction}

How the solar wind interacts with a planet is a major driver of mass, momentum, and energy transport throughout a planetary system.
At planets that have negligible internal magnetic fields, like Venus, the solar wind magnetic field drapes across the dayside of the planet before stretching far downtail from the planet \cite<e.g.>[]{McComas1986, Rong2014}. \note{Rong reference added.}  
In the interaction region around the planet, often called an induced magnetosphere, the solar wind can interact with the upper altitudes of the atmosphere, including the ionosphere \cite{Luhmann1997}. 
On the dayside of the planet magnetized solar wind plasma must flow around the planet.
This deforms and mass loads magnetic fields lines as the plasma diverts around the obstacle \cite{Luhmann1986, Law1995, Dubinin2021}. \note{Dubinin reference added.}
The solar wind interaction on the dayside and in the tail exhibits dawn-dusk asymmetries in magnetic field draping because of the solar wind's inherent magnetic field geometry, \add{often referred to as a Parker spiral angle} \cite{McComas1986, Liemohn2017, Parker1958}. \note{Parker reference added}.
\add{The overarching goal of this work is to determine how much of Mars' solar wind draping interaction can be credited purely to processes in induced magnetospheres.}
\remove{An overarching goal of this work is to determine the extent to which the Martian solar wind draping interaction is the results from processes inherent to induced magnetospheres.}

%(2) Indicate the ways in which Mars exhibits some of the same features.
To a first approximation, Mars' interaction with the solar wind is similar to that of an unmagnetized body where the solar wind magnetic field drapes around the planet and geometries of draping on the dayside propagate downtail \cite{Luhmann1991a, Luhmann1991b, Luhmann1992, Crider2004}. 
This process begins upstream of the bow shock as the solar wind accelerates planetary pickup ions and slows to conserve momentum \cite{Halekas2017}.  
The plasma experiences a sharp deceleration at the bow shock, after which the subsonic plasma continues to slow and divert around the planet, with a geometry that largely depends on upstream solar wind conditions, including within the magnetosheath \cite{Dong2019, Nagy2004}.
Other effects characteristic of unmagnetized planets can be observed at Mars via comparisons to Venus and shared features in their solar wind interaction \cite{Cloutier1999}. 
One significant similarity are hemispheric asymmetries driven by the convection electric field ($\mathrm{E_{sw} = -V_{sw} \times B_{sw}}$) including circumpolar magnetic fields corresponding to the $\mathrm{-E_{sw}}$ hemisphere \cite{Chai2019} and \add{other $\mathrm{-E_{sw}}$ hemisphere dependent asymmetries} \remove{planetary north-south deflected fields} \cite{Zhang2022}. 
However, unlike a purely intrinsic magnetosphere the Mars-solar wind interaction is significantly complicated by the presence of strong crustal magnetic fields \cite<see review within>[]{Nagy2004}.

Crustal magnetic fields at Mars are an important intrinsic feature of the system \cite{Connerney2004}. 
%This is a big of a 'why does it matter' point, but strays the reader
%The physical processes that have resulted in current crustal field observations and when such processes occurred, are still being understood and likely will require additional measurements at lower altitudes than currently available \cite<see>[for more details]{Mittelholz2022}. 
%
Mars' crustal magnetization is non-uniformly distributed across the planet with the strongest field intensities in the southern hemisphere \cite{Langlais2019, Gao2021}. 
These field\add[ver2]{s} vary in strength by up to three orders of magnitudes and rotate with the planet, resulting in a variable obstacle to the solar wind.
Since their discovery \cite{Acuna1999}, crustal fields have been shown to influence the Mars-solar wind interaction, ionospheric structure, and atmospheric loss. 
Evidence from Mars missions have shown variations in magnetic field topology presumed to be a result of reconnection with crustal fields \cite{Xu2020}. 
Similarly, ionospheric composition and ion loss have been shown to vary depending on crustal field strength and location \note{Zhang reference added.} \cite{Franz2006, Fang2017, Flynn2017, Withers2019, Weber2021, Dubinin2020, Zhang2021}.
Crustal fields are thought to lead to departures from the nominal magnetic field draping that would occur for an un-magnetized planet. 
As a result of this complexity, it is difficult to disambiguate internal (e.g. exosphere, ionosphere, crustal magnetic fields) from external (e.g. solar wind) drivers of the magnetic morphology.
This ambiguity itself is increasingly understood as a \add[ver2]{defining}\remove[ver2]{defintional} characteristic of the Mars system with an ongoing interplay of internally, and externally driven features.
This has led to Mars' solar wind draping interaction and region being called a hybrid magnetosphere, which, by definition, exhibits features of an induced interaction and of draping unique to Mars' crustal field effects \cite{DiBraccio2018, DiBraccio2022, Xu2020}.

%(4) Review recent publications in which observations are explained in terms of a hybrid magnetosphere.
Several features in the space environment have been used to characterize Mars as a hybrid magnetosphere. 
For example on the dayside of the planet, observed asymmetries in draping direction have been associated with a preferred direction of the solar wind's interplanetary magnetic field (IMF), possibly through interactions with the crustal field \cite{Brain2006a, Brain2006b}. 
\remove{At higher dayside altitudes departures from simplified draping scenarios have been shown in physical models and associated with the velocity deflection of the solar wind around Mars (Fang et al., 2018).}
On the nightside of the planet, Mars' magnetotail field lobes exhibit a twist away from results expected for a solely ionospheric interaction \cite{DiBraccio2018, DiBraccio2022}.
This twist is suggested to be associated with crustal fields, potentially through magnetic field orientations that would be more or less amenable to reconnection with the incoming solar wind magnetic field.
From these prior studies it is clear that a complex interplay of intrinsic and induced effects are active in influencing the solar wind interaction at Mars.

%(5) Describe how this paper will contribute to what we know so far
In this work, we use magnetic field data around Mars from the Mars Atmosphere and Volatile EvolutioN (MAVEN) mission to assess the solar wind draping configuration at Mars from the dayside to the tail.
Magnetic field morphology is particularly interesting because it influences the loss of plasma from the main ionospheric source region, where ions are cold, controlled by magnetic fields, and initially accelerated \cite<see>[]{Withers2009, Hanley2021}. 
Given the asymmetries observed on the dayside and in the magnetotail, in this study we use MAVEN magnetic field data to globally connect these regions responses under different solar wind conditions. 
We investigate the global draped magnetic field morphology with a \remove{new} statistical field line tracing method. 
This method provides data-driven global insights that have similar resolution to those of physical models.
We then build a Bayesian multiple linear regression model to separate, quantify, and understand the effects of the solar wind, remanent crustal, and induced magnetic fields on dayside magnetic field draping.
Finally, we propose an adjustment to expectations of magnetic field draping in induced magnetospheres to include previously underestimated effects of the full three dimensional solar wind magnetic field geometry.

%----------------------------------------------------------------------------------
\section{Data and Methodology}\label{methods}

We use magnetic field data, estimations of the crustal magnetic field, and of the upstream solar wind magnetic field (Section \ref{proxies}) with field line tracing and a regression model (Section \ref{tools}) to investigate the dependency of solar wind magnetic field draping on the upstream solar wind IMF geometry and remanent crustal fields. 
Our main data are the magnetic field data from MAVEN. 
The MAVEN magnetometer (MAG) investigation consists of two fluxgate magnetometers that deliver rapid (32 Hz) measurements of the three-dimensional magnetic field vector along MAVEN's orbital trajectory \cite{Connerney2015}. %
We use these data at either 1 second (Section \ref{findingsOne}) or 30 second resolution (Section \ref{findingsTwo}, \ref{findingsThree}) between October 2014 and February 2020, ${\sim}11,000$ orbits of the MAVEN spacecraft around Mars.
\add{We use 30 second resolution by sampling the 1 second data at a 30 second cadence.} 
\add{This is used in cases where the full resolution is computationally prohibitive and when better resolution does not affect our results.}
%
%168739200, 10931.0 -> actual data sizes
%
We use magnetic field data in Mars Solar Orbital (MSO) frame where $\hat X$ points from Mars to the Sun, $\hat Z$ points to Mars' ecliptic north, and $\hat Y$ completes the right-handed Cartesian set. $\hat Y$ points opposite to the planet's orbital angular velocity and roughly eastward when viewed on the dayside in the IAU Mars frame.
We then combine these data with estimations of upstream solar wind parameters for analyses in this paper.

\subsection{Estimation of the Interplanetary Magnetic Field (IMF)}\label{proxies}
\remove{Estimates of the IMF}

Because of precession, MAVEN's orbit is in the upstream solar wind only part of the time.
On orbits where MAVEN leaves the Martian magnetosphere and can directly sample the solar wind, we use these direct measurements of the upstream solar wind \cite{Connerney2015, Halekas2015, Halekas2017}.
Where this is not possible, we use a solar wind proxy developed by \citeA{Dong2019} which estimates the upstream conditions from direct measurements in the magnetosheath. 
When neither estimates from direct solar wind measurements or the proxy developed by \citeA{Dong2019} are available we generate the remaining from \citeA{Ruhunusiri2018} which uses an artificial neural network to approximate the solar wind from MAVEN data. 
\add{These formerly developed proxies of the IMF direction are accurate to within 45 degrees for over half of the estimates} \remove{For orbits where both direct and proxy measurements are available they agree to within 45$^{\circ}$ for over half of the estimates from both methods} \cite{Dong2019, Ruhunusiri2018}.
\add{This allows us to use these previously developed proxies to roughly estimate the upstream IMF direction for qualitative conclusions.}

At Mars, the IMF has a dominant magnetic field component, or Parker spiral angle, in the $\hat Y$ MSO direction and secondarily, $\hat X$ MSO direction and is roughly bimodal with two prevalent orientations, $\mathrm{+B_{y}, -B_{x}}$ and $\mathrm{-B_{y}, +B_{x}}$ \cite{Parker1958}.
This can be seen in the IMF cone angle, the magnetic field angle within the X-Y plane and the clock angle, the angle within the Y-Z plane. 
In Section \ref{findingsOne} and \ref{findingsTwo} we use \remove{these} direct measurements and proxies described in the paragraph above of the clock angle to estimate the upstream IMF geometry.
With the use of the proxy measurements, these estimations are often at a more frequent resolution than the solar wind's natural variation \cite<see>[for auto correlation timescales]{Marquette2018}. 
Similarly, while dynamic events such as coronal mass ejections \cite{Lee2018} can perturb this system the results presented here are based on millions of data points in which extreme, but rare, events have little impact. 
In addition to the MAVEN measured magnetic field we include in our analyses: a crustal field model \cite{Langlais2019}, the spatio-temporal location of the MAVEN spacecraft, and topographic elevation from the Mars Orbiter Laser Altimeter onboard the Mars Global Surveyor \cite{Zuber1992} \add{for reference purposes in certain figures}; courtesy of the MAVEN mission team public IDL Toolkit available on the MAVEN Science Data Center.

\subsubsection{Continuous Estimation with Gaussian Processes To Enable Analyses}\label{gpr}

\remove{Estimation of the IMF with Gaussian Processes}

The use of previously derived solar wind proxies is sufficient for large-scale classification (e.g. bimodal solar wind estimations) but for the regression analyses pursued in Section \ref{findingsThree} we need more frequent estimations of the upstream solar wind and their error as current proxy estimations are usually estimated once every unique orbit of MAVEN.
We developed an estimation of the IMF direction $\mathrm{(B_{x}, B_{y} , B_{z})}$ using a Gaussian process regression (GPR).
The GPR is trained on directly measured solar wind values from MAVEN with the original solar wind magnetic field data from MAVEN \cite{Connerney2015, Halekas2015, Halekas2017}.

Gaussian process regression is often employed as a method for interpolation on sparse spatio-temporal data.
This method is a non-parametric\remove{and flexible} machine learning technique that models the relation between inputs and outputs of a multivariate function as a distribution of possible functions. 
GPR is useful in our context because uncertainty is explicitly included in the model formation.
This is advantageous when estimating confidence in predictions \cite<see>[for introductions to Gaussian processes]{Rasmussen2006, Gortler2019}.
Gaussian processes require specification of a prior (kernel) as an initial estimate of the underlying covariance function \cite<see>[for discussion on choosing kernels]{Duvenaud2014}.
We chose to employ GPR for two reasons. 
First, it enables us to generically estimate the IMF during periods when it is not directly measured by MAVEN at effectively infinite time resolution.
Second, along with predicting a mean value, we also use the GPR to estimate a standard deviation that is inversely proportional to our sampling rate. 
In other words, when MAVEN has poor solar wind sampling, our \remove{GPR}\add{IMF} estimate has increased errors.  
We incorporate these error estimations into our subsequent analysis.

When implementing our GPR we tested various kernels before choosing a Rational Quadratic kernel which is an infinite sum of squared exponential kernels of various length-scales \cite{Rasmussen2006}. 
We employed this kernel as it allows for multiple length scales (e.g. variation timescales) which given MAVEN sampling is advantageous.  
We did however cross compare to other possible kernel choices by assessing the regression performance on a test set of data chosen by randomly selecting 10$\%$ of the first 5,000 values. 
We assessed various kernel choices on this test set with the coefficient of determination (R$^{2}$) and by assessing the normality of the residuals. 
We want our predicted values to be normally distributed around the real value (i.e. without systematic errors) which is why we assess the normality of the residuals. 
For the Rational Quadratic kernel our R$^{2}$ values are 0.98, 0.98, and 0.95 for B$_{x}$, B$_{y}$, and B$_{z}$ respectively. 
Ideally a value of 1 is a perfect prediction.
The means of the residuals are 0.05, -0.06, 0.08 (compared to 0 for a perfect normal), and the standard deviations of residuals of 1.04, 1.02, and 0.95 (compared to 1 for a perfect normal). 
This means that our GPRs errors as measured on our test set are roughly distributed normally around our true value. 
We also check the performance visually which is relatively simple as we are only using the GPR in one dimension, time.

Regardless of the high performance of our GPR there are limitations to using any interpolation method on a data set that is sampled irregularly.
For example, our GPR has large estimated errors in regions where the IMF is very sparsely sampled (i.e. multiple sequential orbits with no solar wind measurements).
Because of the method of the GPR we employed, these sparsely sampled times with large errors often \add{trend}\remove{default} to predicting the average value.
It is easy to identify when the GPR fails to predict realistic values by comparing the mean value of the GPR verses the standard deviation.
We find that when the standard deviations are less than $\sim$0.8 nanoTesla (nT) the GPR predictions are smoothly varying between MAVEN measurements and represent a physically feasible prediction of the IMF. 
In our analyses then we only use GPR predicted values when the standard deviation is under 0.8 nT.
This exact standard deviation cut off is flexible and our results in subsequent sections of this paper are not greatly sensitive to this exact value. 

\subsection{Analysis Tools}\label{tools}

We investigate the global response of Mars' magnetic field environment to external and internal factors in two ways. 
First, we use magnetic field line tracing to derive qualitative insights into Mars' hybrid magnetosphere under each of the two primary IMF conditions at Mars $\mathrm{(+B_{y}, -B_{x}}$ and $\mathrm{-B_{y}, +B_{x})}$. 
Second, we use Bayesian multiple linear regression to disambiguate observed effects.

\subsubsection{Statistical Tracing of Magnetic Fieldlines}\label{tracing}

To date, studies of the large-scale magnetic field in Mars' hybrid magnetosphere have often relied on physical models or limited spatial studies. 
To investigate the global nature of Mars magnetic field, we have developed a tracing method that provides insights with similar resolution to that of physical models but is sourced from measured data.
As discussed within \citeA{Liemohn2017}, MSO potentially conflates trends by disregarding geographic coordinates or hemispheres.
We preserve geographic coordinates by tracing magnetic field lines in MSO, but only over locations that can occur concurrently (e.g. 180$^{\circ}$ geographic longitude at noon requires 0$^{\circ}$ geographic longitude at midnight) and under the same average IMF direction. 
We use this method to create average field line maps for a given set of \add[ver2]{sub-}solar longitude and IMF direction.
This allows us to observe dawn-dusk asymmetries, geographic dependencies, and preserve effects that are organized by sub-solar longitude. 
This method provides as close a view to what could be observed if a sensor existed to take simultaneous global observations of Mars' magnetic field under different solar wind conditions.

\begin{figure}[ht]
 \noindent\includegraphics[width=\textwidth]{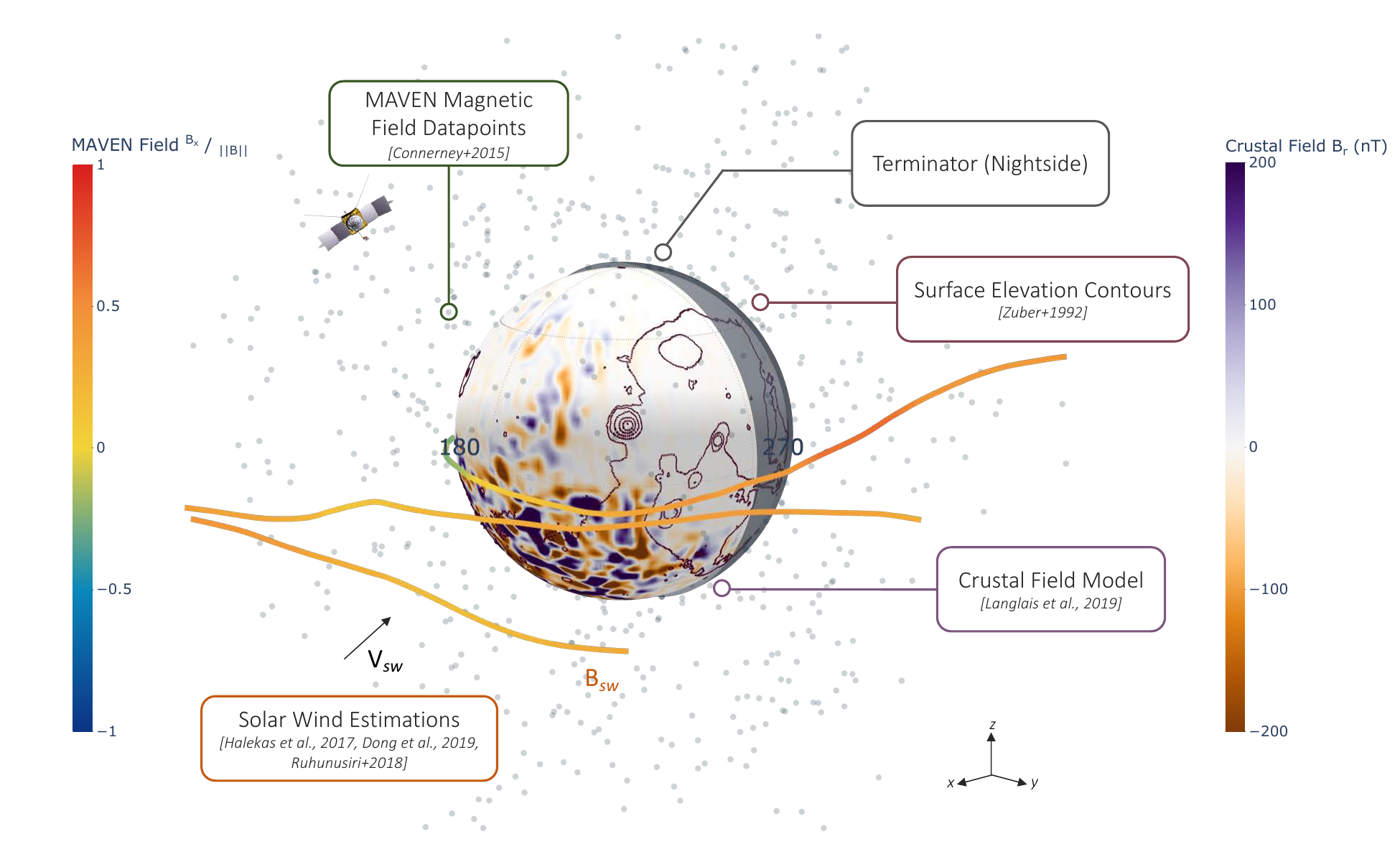}
 \caption{Field line tracing method for a westward $\mathrm{(-B_{y}, +B_{x})}$ solar wind magnetic field when the Sun location aligns with 180$^{\circ}$ geographic longitude. 
 This figure illustrates three tracings of the magnetic field (at 500, 2500, and 4500 km over 0$^{\circ}$ latitude and 180$^{\circ}$ geographic longitude) with the above solar illumination and solar wind magnetic field conditions. 
 The pale grey points mark every 2500$^{th}$ datapoint as observed by MAVEN within 2.5 Mars radii representing the approximately 3 million measurements available under these spatial and IMF conditions with the field-line tracing method pursued. 
 Solar wind estimations in this figure are 55.3$\%$ directly measured by MAVEN \cite{Connerney2015, Halekas2015, Halekas2017}, 44.4$\%$ approximated by measurements in the magnetosheath \cite{Dong2019}, and 0.3$\%$ as estimated from a neural network \cite{Ruhunusiri2018}.
 \add{Elevation contours are provided in dark red calculated from} \citeA{Zuber1992}.
 The traced field lines demonstrated expected solar wind draping as the incoming magnetic field lines deform and drape over the planet.}
 \label{fig:methods}
 \end{figure}

In Figure \ref{fig:methods} we demonstrate this field line tracing method for a subset of traces. 
This figure is of a westward $\mathrm{(-B_{y}, +B_{x})}$ IMF when the sub-solar longitude is 180$^{\circ}$. 
\add{We pursue analyses with this method focusing on $\mathrm{(+B_{y}, -B_{x})}$ and $\mathrm{(-B_{y}, +B_{x})}$ orientations.}
\add{More specifically we take all the data that occurred under +Y or -Y estimated IMF direction within a $\pm$ 45$^{\circ}$ clock angle.}
\remove{We pursue analyses with this method focusing on $\mathrm{(+B_{y}, -B_{x})}$ and $\mathrm{(-B_{y}, +B_{x})}$ orientations determined by restricting data to a $\pm$ Y estimated IMF direction within $\pm$ 45$^{\circ}$ clock angle.}
We use these two IMF orientations for four sub-solar locations, when the Sun is aligned with 0$^{\circ}$, 90$^{\circ}$, 180$^{\circ}$, and 270$^{\circ}$ geographic longitude.
We bin by 15$^\circ$ planetary longitude and 1 hour local time \add{which is the minimum binsize to obtain robust results}. 
For example, plots which show 0$^{\circ}$ longitude at local time noon have MAVEN data sampled between: 0-15$^\circ$ longitude and 12-13 local time, 15-30$^\circ$ longitude and 13-14 local time, and so on progressing around the planet.

Magnetic field traces are calculated for 60 points initiated at 500 km altitude distributed globally on a Fibonacci lattice. 
Two additional traces are calculated at 2500 and 4500 km at the equator and local time noon which are used to demonstrate the IMF direction. 
We use a Fibonacci lattice as the starting place for each of the 60 traces at 500 km as this approximates an even sampling on the surface of a sphere with each original point representing approximately the same area \cite<see>[for details]{Swinbank2006, Gonzalez2010}.
This is a desirable property for comparing polar and equatorial observations and avoids issues of non-equal latitude by longitude grid spacing.

Traces are computed as a propagation through a series of averages where each subsequent step in the traced field line is calculated from the previous average magnetic field direction. 
To \add{internally} validate this method, we tested it on a dipole magnetic field with data sampling conditions identical to the MAVEN dataset. 
We find the technique \add{qualitatively} reproduces what is expected from a dipole field and is resilient to spatial data sparsity.

%
%Unlike physical models, there is no restriction on these field lines to adhere to solutions to Maxwell's equations and instead traces capture the average behavior of the measured field under specified IMF directions and planetary orientations.

\subsubsection{Bayesian Multiple Linear Regression} \label{bayesianregression}

To determine the dependence in our observations on crustal fields and the IMF we have implemented a Bayesian multiple linear regression.
Bayesian methods generally follow Bayes rule in order to estimate confidence in outcomes given certain data\add{, or observations} \cite<see>[for original discussion]{Bayes1763}.
\add{While data science methods have not been as readily employed in planetary science as compared to other fields, partially due to the sparsity of planetary data,} \cite<see>[]{Azari2021} \add{they are widely applicable}.
\add{Bayesian methods, including linear regression,  have become commonly employed for statistical analysis, particularly within astronomy and biology} \cite<e.g.>[]{Thrane2019, Wilkinson2007}.
\add{We direct the reader to} \citeA{Kruschke2015} 
\add{for a general introduction to Bayesian methods for data analysis.}
\add{In short, the method employed within this work, Bayesian multiple linear regression, describes the outcome of a variable as a linear combination of other variables.}
\add{Often this method is employed, as we do here, to determine the predicted distribution of regression coefficients to understand, or infer, some knowledge of a physical system.}

\add{In this work we wish to quantify, with error, the dependence of our findings on multiple potential physical parameters.}
\add{This need has led us to employ a Bayesian regression as it enables uncertainty incorporation from the solar wind IMF and allows estimation of both intrinsic, induced, and direct solar wind effects.}
\remove{Our use of the GPR in Section 2.1.1 is a Bayesian approach.}
\remove{The main reason we implemented a Bayesian regression is to incorporate the uncertainty of the solar wind IMF into our interpretations.}

The linear model we developed is described in equation \ref{eqn:model} in which we model the measured magnetic field over the dayside (10 - 14 hours local time, 15 - 75$^{\circ}$ MSO latitude) as a linear combination of the IMF and crustal fields.  
We use this model to compare the $\beta$ coefficients to estimate the importance of each physical effect.
\begin{linenomath*}
\begin{eqnarray} \label{eqn:model}
 \mathrm{B^{\star}_z,_{MAVEN} = \beta_{0, Intrinsic} \cdot B^{\star}_z,_{Crustal} \; +} \nonumber \\
 \mathrm{\beta_{1, Solar Wind} \cdot B^{\star}_z,_{GPR} \; -} \nonumber \\
 \mathrm{\beta_{2, Induced} \cdot B^{\star}_x,_{GPR} \cdot sign(Z_{MSO}) \; +} \nonumber \\
 \mathrm{\beta_{3, Intercept} + \epsilon}
\end{eqnarray}
\end{linenomath*}
In words, we expect that MAVEN measured magnetic field in $\hat Z$ to be a linear combination of: the crustal field \cite{Langlais2019}, the solar wind IMF in $\hat Z$ estimated from our GPR (Section \ref{gpr}), an induced effect which depends on the solar wind IMF in $\hat X$ (also estimated from our GPR) with a sign dependence on hemisphere, and a general intercept \add{(analogous to a y-intercept)}. 
\add{The third term is negative as an aesthetic choice instead of using $\mathrm{-1*sign(Z_{MSO})}$}. 
The last term $\epsilon$ represents  the intrinsic uncertainty, or scatter around our chosen model, which we constrain to follow a centered Gaussian distribution with variance $\sigma^2$ in accordance with the assumptions of the underlying regression model. 
We report $\sigma$ in our results as an estimate of intrinsic scatter in this model which can be considered as a variation around any identified trends.
\add{Our goal in this fit, as mentioned previously, is to estimate the relative importance of the physical parameters included here by comparing the scale of the coefficients  $\beta_{i}$}.
\add{For example if our estimated distribution of $\beta_{0},_{\mathrm{Intrinsic}}$ is larger than $\beta_{1},_{\mathrm{SolarWind}}$ then we would conclude that the crustal fields are playing a larger role in describing our observations.} 
\add{Similarly, if $\beta_{3},_{\mathrm{Intercept}}$ is large it can be considered as a hint that there are other variables that result in an offset in our observations that are not captured in this model.}

\add{This coefficient comparison is valid for two reasons.} 
\add{First, all data used in Equation} \ref{eqn:model} \add{are normalized by subtracting the mean and dividing by the standard deviation which is considered a standard method (normalized quantities are denoted with $^{\star}$).} 
\add{Second we are using a Bayesian interpretation of ridge regression which regularizes the scale of our coefficients.}
\add[ver2]{Bayesian methods require setting a prior, or an original hypothesis about distributions. For linear regression we need to set priors to describe the distribution of coefficients.}
\add{We use Gaussian priors for the coefficients centered on zero with a standard deviation of 5 after testing an array of standard deviations from $1$, for a strictly regularized fit, to $10$, for a weakly regularized fit.}
\add{We find our results to be the same within tolerance regardless of this choice, indicating that the quality of the signal within the data is strong enough to mitigate the need for a strong prior} \cite<see>[for discussion on prior choices]{James2013}.
\remove{We perform the regression using Gaussian priors on the coefficients centered on zero with a standard deviation of 5 after testing an array of standard deviations from $1$, for a strictly regularized fit, to $10$, for a weakly regularized fit.
We find our results to be the same within tolerance regardless of this choice, indicating that the quality of the signal within the data is strong enough to mitigate the need for a strong prior (see James et al., 2013 for discussion on prior choices). 
All data used in Equation 1 are normalized by subtracting the mean and dividing by the standard deviation which we denote with $^{\star}$ and are used in MSO.}
GPR estimates of the IMF are included with their predicted standard deviations in this regression following the process detailed in the supplementary software within  \citeA{Abrahams2022}. % make clear that the ooutput of the GPR is what is being include, output of the GP, make it clear it's not simultaneously solved
The physical motivation of the variables in our model is described in Section \ref{findingsThree}. 

%----------------------------------------------------------------------------------
\section{Results}

\subsection{Global Morphology of the Draped Solar Wind Magnetic Field} \label{findingsOne}

\begin{sidewaysfigure}
 \noindent\includegraphics[width=\textwidth]%{./FigFrontFacing.pdf}
 {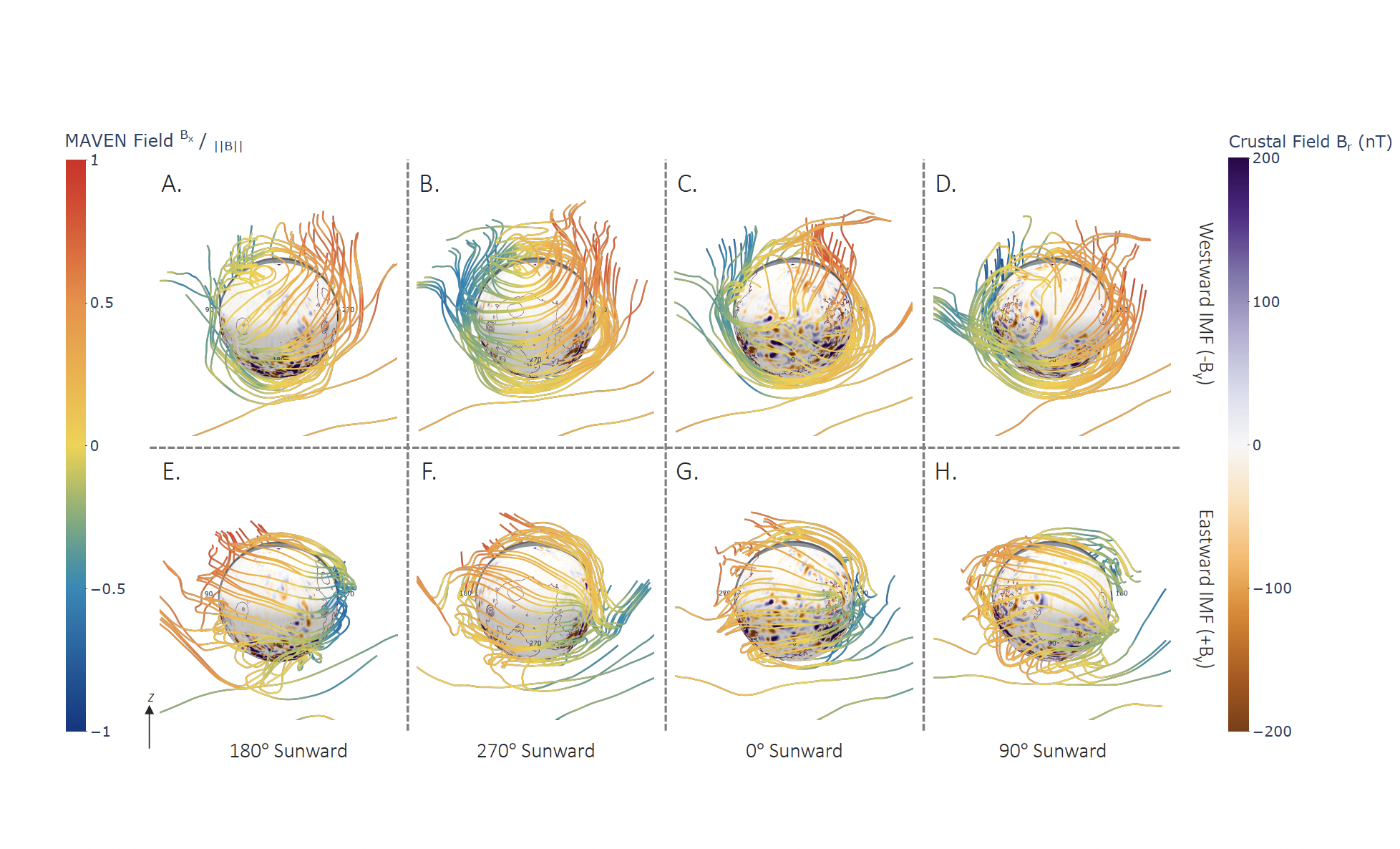}
 \caption{Average magnetic field morphology under different solar wind IMF conditions and Sun-Mars angles. 
 Panels A - D detail the draped magnetic field under westward IMF conditions (-B$\mathrm{_{y}}$, +B$\mathrm{_{x}}$) and panels E - H detail similarly under eastward IMF conditions (+B$\mathrm{_{y}}$, -B$\mathrm{_{x}}$). 
60 traces are originated at 500 km and two traces show the upstream IMF conditions traced at 2500 and 4500 km by the process detailed in Section \ref{tracing}
 This figure uses the same color bars as shown in Figure \ref{fig:methods} with red shaded field lines pointing toward the Sun and blue shaded lines pointed away from the Sun. 
 The crustal field is calculated from \citeA{Langlais2019}.
 \add{Elevation contours are provided in dark red calculated from} \citeA{Zuber1992}.}
 \label{fig:dayside}
 \end{sidewaysfigure}

We utilize our field line tracing method to investigate the magnetic field morphology over Mars. 
In Figure \ref{fig:dayside} we present the average draping over the planet at 500 km altitude for different sub-solar longitudes.
By comparing the top and bottom row in Figure \ref{fig:dayside} it is clear that the magnetic field morphology of Mars under eastward and westward IMF conditions differs significantly.

Under westward $\mathrm{(-B_{y}, +B_{x})}$ conditions traced field lines exhibit a circular pattern over the northern regions of Mars. 
This reverses to the southern hemisphere (not shown) for eastward $\mathrm{(+B_{y}, -B_{x})}$ IMF such that the northern hemisphere no long exhibits these circling field lines. 
While the crustal field appears to inhibit this trend (e.g. panels C and D do not show circular fields, whereas panels A and B do) this circling trend is coherent pattern that persists across several crustal field-Sun line orientations.
The circumpolar fields in panels A and B over the northern hemisphere for a westward IMF correspond with the -E$\mathrm{_{ V \times B}}$ hemisphere. 
It has been proposed that these circumpolar fields are a result of solar wind ions (H$^{+}$) moving in the opposite direction of the planetary plume (O$_{2}^{+}$, O$^{+}$) to conserve momentum drag field lines toward toward the $\mathrm{-E}$ hemisphere \cite<see discussion in>{Dubinin2018, Chai2019}. 
This is an ubiquitous feature present in the Mars system resulting in the large scale upwards or downwards dragging toward the $\mathrm{-E}$ hemisphere \cite<see further global discussion in>{Zhang2022}.
As our method preserves the north-south dichotomy of the crustal fields, we are able to observe how this effect shows hemispheric dependency under eastward and westward IMF conditions.

The physical explanation behind circumpolar fields (H$^{+}$ motion opposite to the pickup plume to conserve momentum) however does not explain the observed twist of the draped magnetic fields over the sunlit hemisphere seen in Figure \ref{fig:dayside}.
For a perfectly induced magnetosphere we would expect draped fields line to stretch horizontally across the dayside of the planet. 
This would result in field lines that have a near zero $\hat Z$ measured field and appear aligned with the planetary latitude \add{with small localized perturbations due to weathervaning effects} \cite{Liemohn2017, Law1995, Dubinin2021}. \note{Dubinin reference added.}  
Instead we observe for a westward IMF a twist that results in a rotated field lines over the \add{entire} dayside.
This is seen in draped field lines in the northern hemisphere diagonally stretching from high latitudes in the dusk hemisphere (right side) to low latitudes in the dawn hemisphere (left side) in Figure \ref{fig:dayside}. 
This pattern appears to reverse for an eastward IMF with field lines diagonally stretching from high latitudes in the dawn hemisphere (left side) to low latitudes in the dusk hemisphere (right side).  
In both cases however, this can be attributed to a $-\hat Z$ directed component of the draped magnetic field in the northern hemisphere.
These results agree with previously studies that showed this effect over isolated regions \cite{Brain2006a, Brain2006b}. 

\begin{sidewaysfigure}
\noindent\includegraphics[width=\textwidth]%{./FigSideFacing.pdf}
{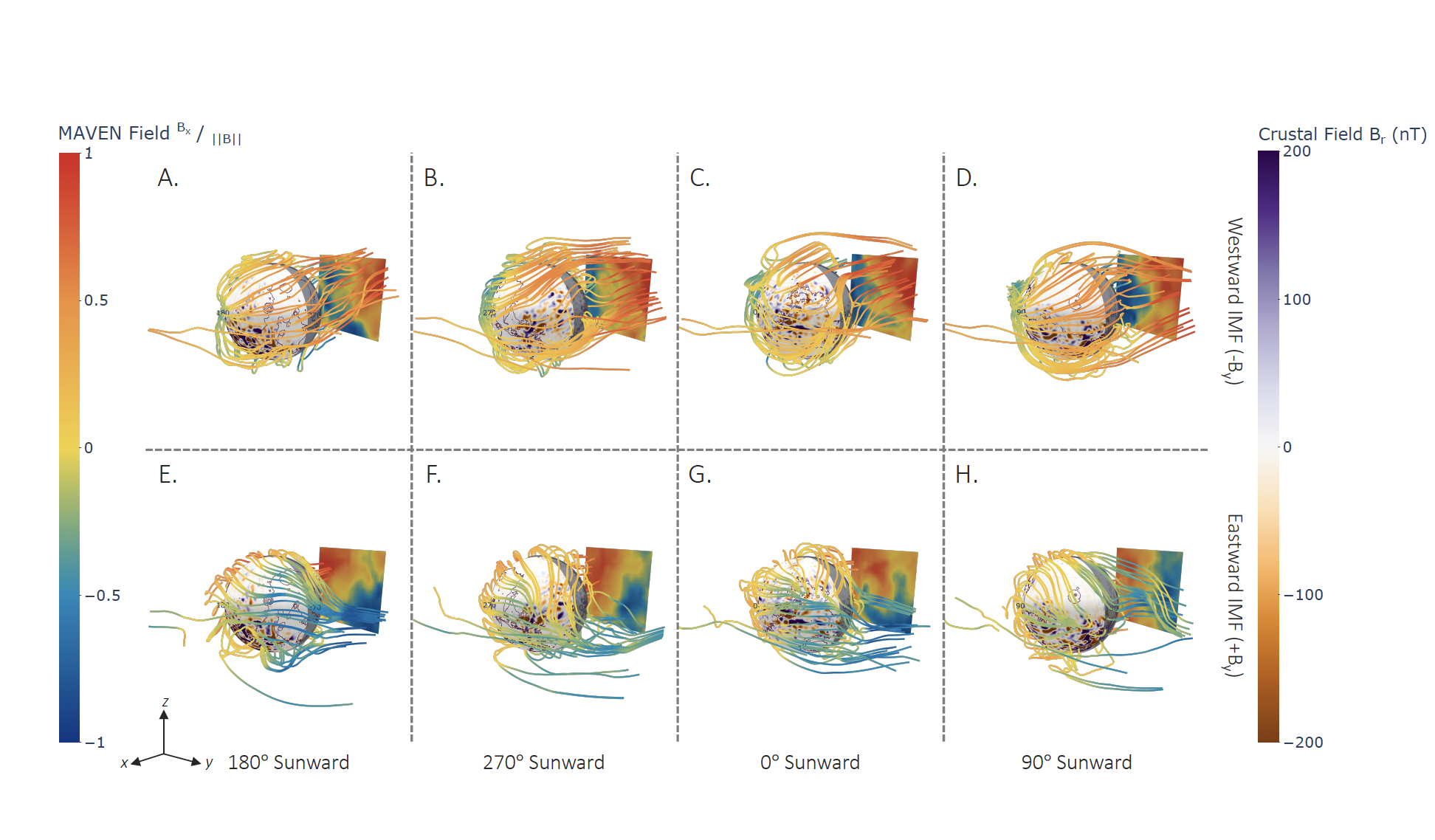}
\caption{Average magnetic field morphology under different solar wind IMF conditions and Sun-Mars angles as viewed from the side. 
Panels A - D detail the draped magnetic field under westward IMF conditions $(\mathrm{-B_{y}, +B_{x})}$ and panels E - H detail similarly under eastward IMF conditions $(\mathrm{+B_{y}, -B_{x})}$. 
60 traces are originated at 500 km and two traces show the upstream IMF conditions traced at 2500 and 4500 km by the process detailed in Section \ref{tracing}.
This figure uses the same color bars as shown in Figures \ref{fig:methods} and \ref{fig:dayside} with red shaded field lines pointing toward the Sun and blue shaded lines pointed away from the Sun. 
This figure additionally shows the average magnetotail magnetic field in $\mathrm B_{x}$ MSO. 
The crustal field is calculated from \citeA{Langlais2019}.
\add{Elevation contours are provided in dark red calculated from} \citeA{Zuber1992}.
} %Panels A and E are included in the Supplementary Material as interactive graphics.
 \label{fig:side}
 \end{sidewaysfigure}

These findings demonstrate a draped field that is distorted along the $\hat Z$ direction over the dayside that is large scale and persists across multiple sub-solar crustal field orientations (Figure \ref{fig:dayside}). 
The magnetotail of Mars has been observed to exhibit rotations in the current sheet location with respect to the X-Z MSO plane that align with the direction of our observations of the dayside twist \cite{DiBraccio2018, DiBraccio2022}.
This can be seen qualitatively in Figure \ref{fig:side} which follows the same method as in Figure \ref{fig:dayside} but from a side view and with the average $\mathrm{B_{x}}$ component shown in the magnetotail.

Figure \ref{fig:side} demonstrates through statistical field line tracing how draped field lines propagates through the terminator region to produce the twisted (shown in $\mathrm B_{x}$ in MSO) magnetotail current sheet orientation.
Both the dayside draped fields, and the current sheet location exhibit the same orientation.
The downwards $-\hat Z$ deflected fields on the dayside result in a counterclockwise rotation of the magnetotail current sheet for a westward ($\mathrm{-B_y}$) IMF and a clockwise rotation of the current sheet for an eastward IMF ($\mathrm{+B_y}$). 
Our results demonstrate that this inner tail twist extends into and likely originates on the dayside.

If the distortion of the dayside draped fields relied heavily on the specific geographic crustal field orientation, we would expect differences in the columns of Figures \ref{fig:dayside} and \ref{fig:side}, but instead we see a roughly consistent draping pattern and inner tail current sheet location.
This suggests that distortions in dayside draping are driven by factors that do not depend on subsolar orientation of the crustal fields.
This could potentially include an as yet underestimated induced feature of the solar wind interaction with the planet or a global (e.g. non-localized) contribution from the crustal fields.

\subsection{Quantifying The Dayside Twist Response to the IMF Cone Angle} \label{findingsTwo} 

The two potential causes of the dayside draping distortions, an underestimated induced feature or a global contribution from the crustal fields, are likely altitude dependent. 
We would expect any solar wind interaction that results in an induced response to be more prevalent at high altitudes, whereas a crustal field contribution should be more prevalent at low altitudes. 
We can investigate this possible dichotomy by focusing on the distribution of the magnetic field by altitude and by IMF direction.

\begin{figure}[ht]
 \begin{center}
 \noindent\includegraphics[width=1\textwidth]{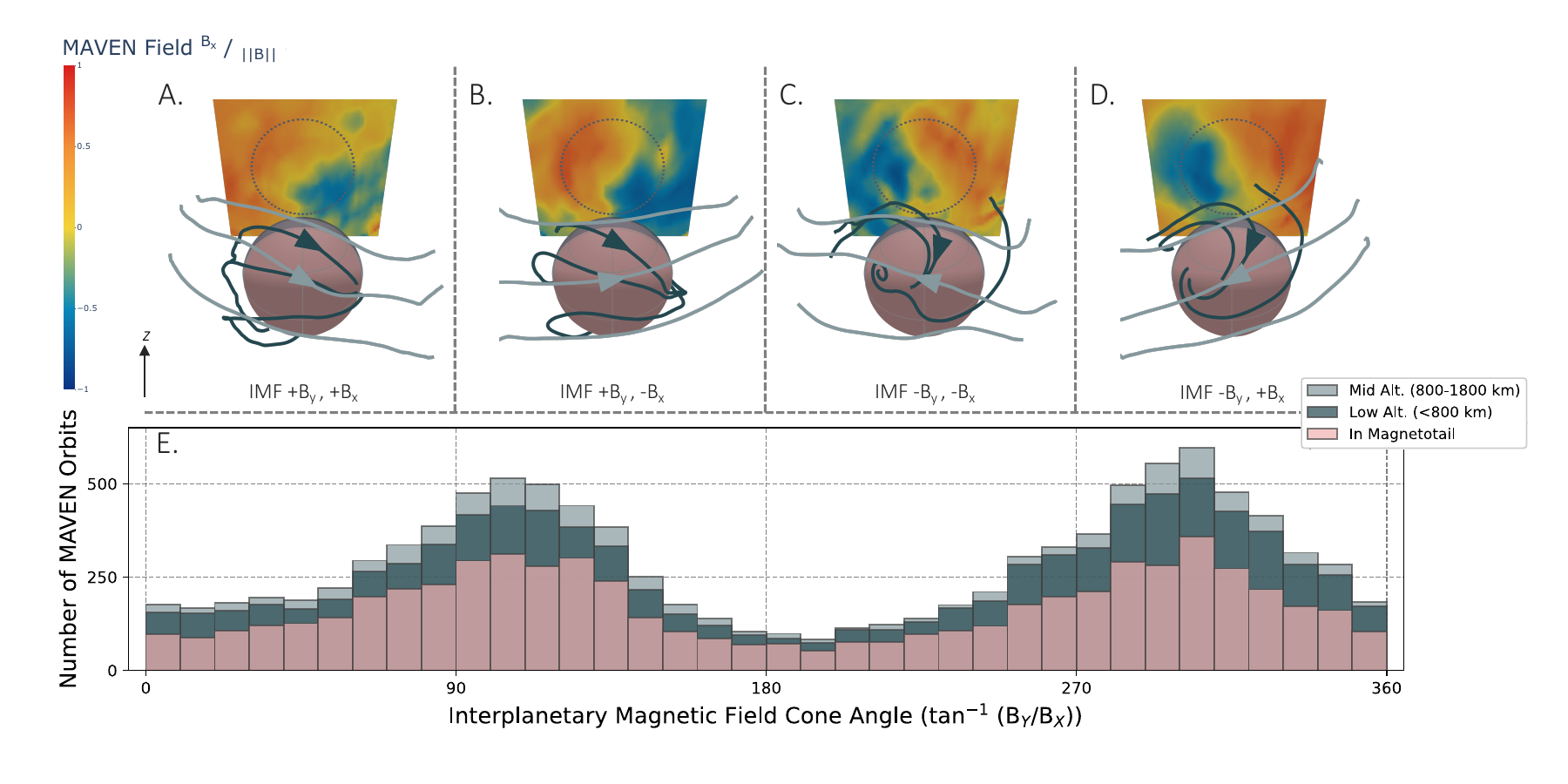}
 \end{center}
 \caption{Altitude dependent dayside magnetic field tracing under variations in the IMF cone angle. 
 Panels A - D show tracings in high (800 - 1800 km, light green) and low (under 800 km, dark green) altitudes regions over the northern hemisphere before showing the average magnetotail magnetic field in $\mathrm B_{x}$ MSO. 
 Arrows on the field line traces demonstrate the field line direction.
 The tail panel in this figure includes a dashed circle marking the 800 km altitude region projected into the tail. 
 This is a rough estimate of where high altitude dayside fields should drape (outside of the dashed circle) and where low altitude dayside fields should drape (inside of the dashed circle). 
 Panel E shows the MAVEN sampling of each IMF cone angle condition.}
 \label{fig:altitude}
 \end{figure}

In Figure \ref{fig:altitude} we trace the magnetic field for two altitude ranges over the the planet and show the magnetotail current sheet orientation (plotted as the average $\mathrm B_{x}$ in MSO) for different IMF directions in the X-Y plane (cone angle).
From this figure it is evident that the magnetic fields at high altitudes ($>$ 800 km, light green lines) are responding to changes in the IMF $\hat X$ component. 
This can be seen by seeing that panels that vary in IMF $\hat X$ direction (e.g. panel A vs panel B, panel C vs panel D), have different high altitude draping directions (arrows on traced field lines).

This high altitude draping response to the IMF $\hat X$ component maps to the outer tail (region outside the dashed circle in magnetotail).
In the magnetotail, the outer region current sheet boundary rotates in the same direction as the upper altitude dayside traces.
This results in outer tail current sheet differences based on the IMF $\hat X$ component (e.g. compare outer magnetotail of panel A vs panel B, panel C vs panel D)

In contrast low altitude traces ($<$ 800 km, dark green lines) all have a -$\hat Z$ directed component on the dayside regardless of the IMF direction (e.g. cross compare panels A through D). 
Similar to the high altitude dayside mapping to the outer tail, we show that the low altitude draping maps to the inner tail.
Because the lower altitude draping is consistent with respect to the IMF $\hat X$ component, the inner current sheet tail location is consistent between panels A vs B, and panels C vs D.

These trends support the hypothesis that higher altitudes are responding to a distinctly different process, possibly an induced solar wind response, whereas lower altitudes are responding to an intrinsic, or crustal field driven process.
It is worth noting that this dichotomy of driving processes by altitude results in differences between an outer and inner tail morphology. 
Given the altitude dependencies that are shown here, it is likely that these differences are contributing to  current sheet morphology differences between the inner and outer tail under different IMF conditions \cite<see>[]{DiBraccio2018}.

\begin{figure}[ht]
 \begin{center}
 \noindent\includegraphics[width=1\textwidth]{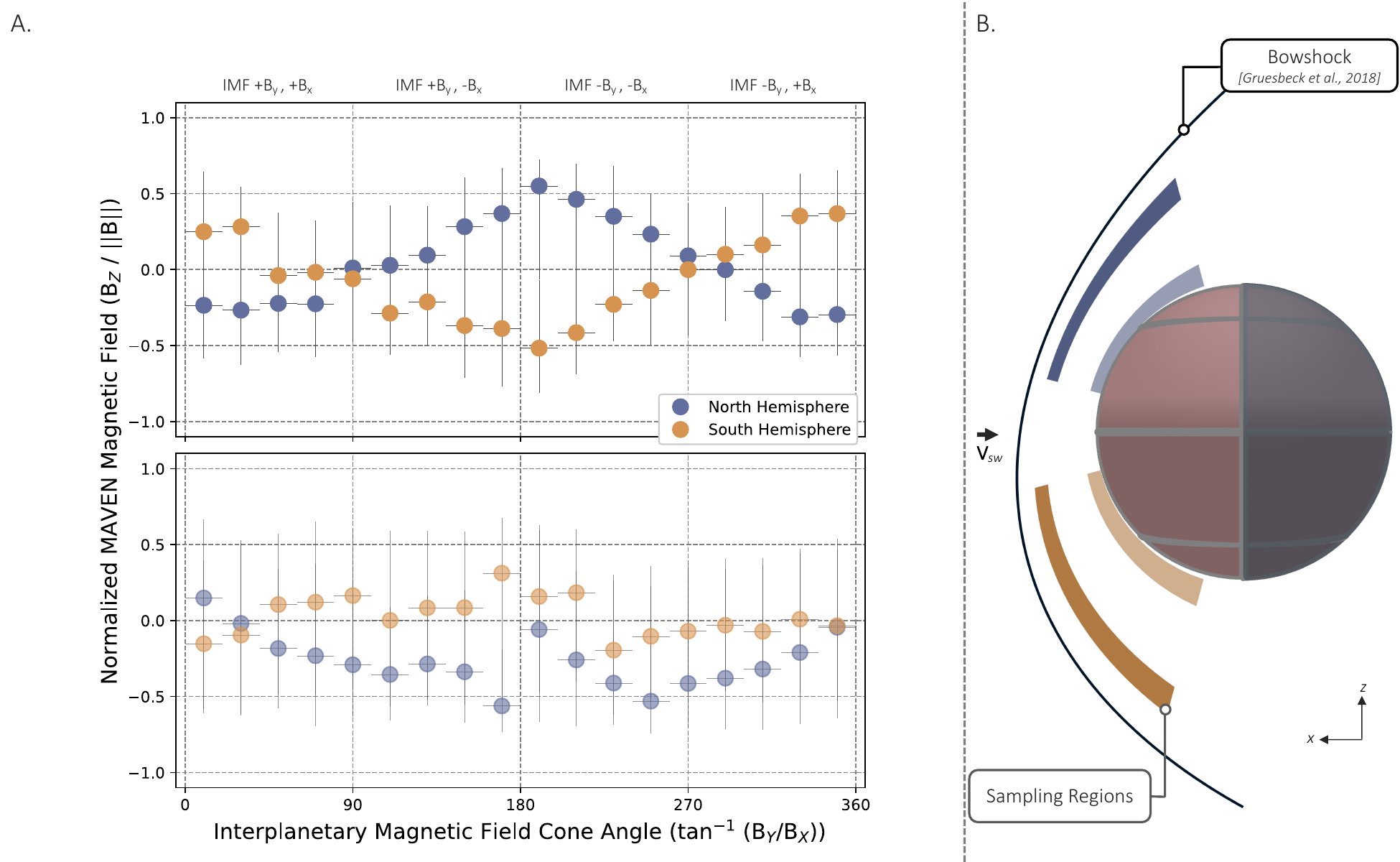}
 %{./FigDaysideTwist_reduced.png}
 \end{center}
 \caption{Dayside magnetic fields twist under variations in the IMF cone angle. 
 A shows the median and interquartile range of the $\hat Z$ MSO directed field versus the IMF cone angle, $\tan^{-1} \mathrm {(B_{y} / B_{x}}$). 
 The top panel of A represents data from 65-80$\%$, and the lower panel of A represents data from 5-20$\%$, of the bow shock distance as calculated from \citeA{Gruesbeck2018}. 
 This is $\sim$100-500 km and 1200-1500 km altitude at noon local time and 0$^{\circ}$ latitude in MSO.
 Panel B shows the sampling region over which these data are taken from 10 to 14 local solar time and 15 to 75$^{\circ}$ MSO latitude.}
 \label{fig:histogram}
 \end{figure}

In Section \ref{findingsOne}, dayside draping distortions in the $\hat Z$ direction resulted in distortions in our expected idealized draping scenario by creating a twisted dayside field which propagated downtail.
To understand what causes the dayside twist and how this propagates to the tail, we quantify the magnetic field $\hat Z$ component in Figure \ref{fig:histogram}. 
We calculate the average $\hat Z$ component over two regions of the dayside, a high alitude region (Figure \ref{fig:histogram} top panel, 65-80$\%$ of the \citeA{Gruesbeck2018} calculated bow shock) and a low altitude region (Figure \ref{fig:histogram} lower panel, 5-15$\%$ of the \citeA{Gruesbeck2018} calculated bow shock). 
At high altitudes, it can be seen that when the IMF is directed toward the planet ($-\hat X$, 180$^{\circ}$ cone angle), magnetic fields distort upwards (+$\hat Z$) in the northern hemisphere, and downwards ($-\hat Z$) in the southern hemisphere. 
When the IMF is directed away from the planet ($+\hat X$, 0$^{\circ}$ cone angle), magnetic fields instead distorts downward ($-\hat Z$) in the northern hemisphere, and upwards ($+\hat Z$) in the southern hemisphere. 
This trend is non-existent at lower altitudes and instead we observe a consistent hemispheric dichotomy with the northern hemisphere containing -$\hat Z$ directed fields. 
Given the lack of dependence of lower altitude $\hat Z$ components on the IMF, this is likely a direct measurement of the crustal fields.
\add[ver2]{Our findings do not show what potential physical mechanisms could lead to dayside draping propagation downtail from these low altitude distortions.}
\add[ver2]{They do show however that the low altitude dayside draping is likely responding to low order moments in the crustal fields and that these responses align with distortions in the tail.}
\add[ver2]{One mechanism that has been proposed is reconnection} \cite{DiBraccio2018, DiBraccio2022} \add[ver2]{but how this could result in allowing transference of the dayside crustal field orientation to the nightside is an open discussion.}   
\remove[ver2]{This doesn't rule out that reconnection is playing a role by altering overall draping patterns or allowing transference of the dayside crustal field orientation to the nightside, but we find in this work that we can credit the dayside and tail draping morphology directly to the crustal field morphology itself.}

\add{Further supporting this exact mapping we show that the northern hemisphere at low altitudes has a roughly consistent -$\hat Z$ component (which we can also see in Figures} \ref{fig:altitude}, \ref{fig:histogram})  \add{and the southern hemisphere at low altitudes shows a roughly consistent +$\hat Z$ component}.
\add{This would result in a S shaped inner tail twist and this is indeed what we observe in Figure} \ref{fig:altitude}. 
\add{This similar morphology was observed by} \cite{Ramstad2020} \add{but not quantified in} \cite{DiBraccio2018, DiBraccio2022} \add{which focused on characterizing a singular twist.} 
\add{Further work is needed to connect these previous studies to the updated understanding within this work on the inner tail.}
\add{At the time of this writing however, the outer tail's response is of greater implication to other induced magnetospheres as this appears to be a direct repercussion of the solar wind interaction rather than any intrinsic difference of the planet.}

Often in idealized draping scenarios at Mars we assume that the $\hat X$ component of the IMF is unimportant unless studying radially directed IMF \cite{Fowler2022}.
However, conditions where $\hat X$ is a major component of the IMF occur frequently at Mars from the typical Parker spiral angle \cite{Parker1958}.
Instead, we find that at high altitudes the $\hat X$ component of the IMF affects the draping pattern over the dayside, leading to twists that propagate downtail. 
We propose that this trend can be explained by considering Mars as a conductive obstacle.
The IMF can not pass through the planet, and instead must deform around it, regardless of its original direction.
\remove[ver2]{This results in the observed deflection at high altitudes  seen observational here and that have previously been captured in magnetohydrodynamics models of Mars}\add[vers]{This results in the deflection at high altitudes observed within this work. This has previously been captured in magnetohydrodynamics models of Mars }\add{and associated with the velocity deflection of the solar wind around Mars} \cite<Figure 1 within>[]{Fang2018}.
This aspect of draping over a conducting obstacle has generally been underestimated. 
As a result, it is unknown how much this affects the dayside draping pattern as compared to the crustal field, or the direct (as compared to induced) solar wind effect. 
In the following section we quantify the relative contributions of each of these effects.

\subsection{Comparison of the Induced and Intrinsic Magnetic Fields at Mars} \label{findingsThree} 

We calculate the contribution of intrinsic and induced magnetic fields to dayside draping by building a Bayesian multiple linear model of $\mathrm{B^{\star}_{z,_{MAVEN}}}$, the normalized $\hat Z$ component in MSO of the MAVEN measured field. 
We expect that the measured dayside $\hat Z$ magnetic field to be a combination of the crustal field in the $\hat Z$ and the IMF in $\hat Z$.
As shown within Section \ref{findingsTwo} there is also an induced effect in which the IMF in $\hat X$ distorts the draped field in $\hat Z$.
We've shown that this effects distorts draped fields over the southern hemisphere in the opposite direction as those over the northern hemisphere (Figure \ref{fig:histogram}). 
This induced deflection is sign dependent on the hemisphere of the planet and so in our model we have included a correction that reverses induced trends in the southern hemisphere by calculating the sign of the MAVEN position in $\hat Z$ MSO.
In summary, our model focuses on quantifying contributions from the intrinsic crustal field as predicted from \citeA{Langlais2019} $(\mathrm{B^{\star}_z,_{Crustal}})$, the direct IMF $(\mathrm{B^{\star}_z,_{GPR}})$, and the induced IMF effect $(\mathrm{B^{\star}_x,_{GPR} \cdot sign(Z_{MSO})})$.
IMF values are estimated from the method described within Section \ref{gpr}.
The full model form and details on this method are defined within Equation \ref{eqn:model}, Section \ref{bayesianregression}.

\begin{figure}[ht]
 \begin{center}
 \noindent\includegraphics[width=0.95\textwidth]{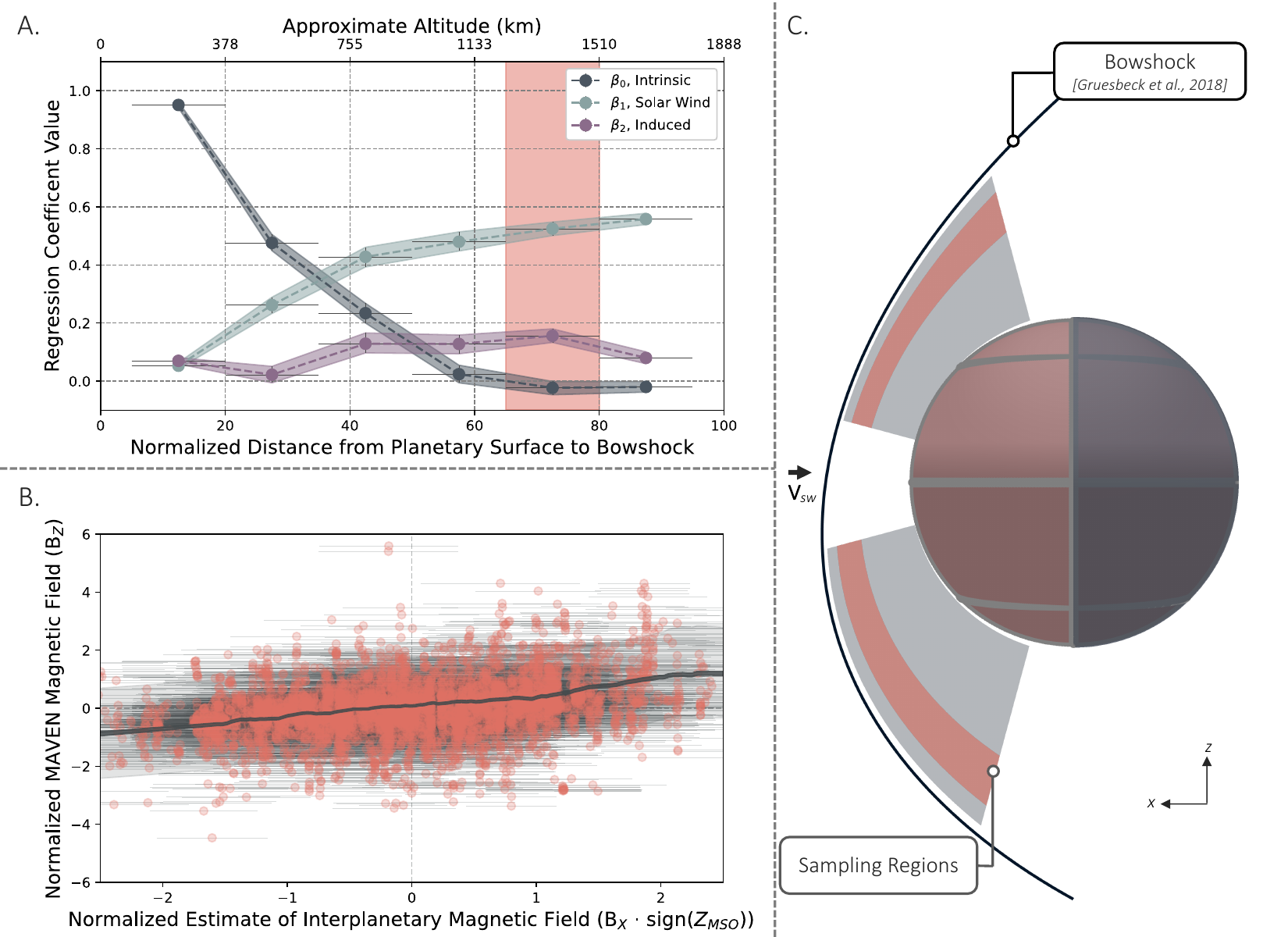}
 \end{center}
 \caption{Estimation of the influence of induced and intrinsic fields on dayside draping.
 Panel A plots the fit coefficients from equation \ref{eqn:model} and their contribution to the observed data by distance to the bow shock from \citeA{Gruesbeck2018} (lower axis) or altitude at noon local time and 0$^{\circ}$ latitude in MSO (upper axis).
 Panel B shows 65 - 80$\%$ bow shock range fit over the intrinsic parameter along with our mean outcome (black  line) and the 95$\%$ credible interval shaded grey.
 Errors on the IMF parameters are calculated from a Gaussian Process Regression (see Section \ref{bayesianregression}), are shown as grey lines, and included in this fit and final estimations of our regression. 
 Panel C shows the sampling region for panels A and B.
 Data within this fit are taking from 10 - 14 local solar time between 15 and 75$\%$ MSO latitude.
 \add{$\beta_{3},_{\mathrm{Intercept}}$ is not shown on this figure as it is measured near a mean value of 0 at all altitudes.}}
 \label{fig:linearfit}
 \end{figure}

We run this regression over the dayside magnetosphere ranging from 5 to 95 $\%$ of the distance between the planet's surface and the bow shock from the model in \citeA{Gruesbeck2018} and then compare the fit coefficients ($\beta_{i}$) from Equation \ref{eqn:model}.
\remove{We can compare the coefficients because we used a Bayesian interpretation of ridge regression on our priors and normalized our dataset.}\note{This is discussed in the methods section now and is redundant.}
Figure \ref{fig:linearfit} panel A shows the values of the fit coefficients at different altitudes. 
It is not unexpected that the dayside draped magnetic fields are primarily predicted by the direct solar wind at high altitudes and the crustal field at low altitudes.
However, we can also see from this panel that the induced effect ($\mathrm{\beta_{2, Induced}}$) is non negligible and larger than the crustal field contribution between 65 and 80$\%$ of the bow shock distance (pink shaded region).
In Figure \ref{fig:linearfit} panel B we plot the normalized MAVEN measured data in $\hat Z$ versus the parameter representing the induced effect $(\mathrm{B^{\star}_x,_{GPR} \cdot sign(Z_{MSO})})$ over this region.
The grey shaded envelope represents a posterior predictive check at the 95$\%$ credible interval and the solid colored line the mean posterior predictive check. 
This check allows to determine if the model encapsulates the \remove{intrinsic} \add{natural} variation of our dataset; 95$\%$ of our data should fall within grey shaded region.

In Table 1 we show the estimated values of the fit coefficients (${\beta_{i}}$) for the highlighted region in pink from Figure \ref{fig:linearfit} (65 - 80$\%$ bow shock range).
From the coefficients within Table 1 it is clear that the $\hat Z$ MSO measured magnetic field over the dayside is well described first by the direct solar wind but secondarily by the induced effect discussed above.
The induced contribution is non-negligible ($\mathrm{\beta_{2, Induced}}$ is above 0). 
We interpret this to mean that direct and induced solar wind effects are largely driving the observed twists in magnetic field at high altitudes. 
Conversely by 20 - 35$\%$ of the bow shock distance (roughly 800 km altitude), Mars' remanent crustal fields are the primary cause of any distorted or twisted dayside fields, and are likely propagating downtail as shown in Figures \ref{fig:side} and \ref{fig:altitude}.

The transition where the crustal field influence, as measured by $\mathrm{\beta_{0}}$, overtakes the direct ($\mathrm{\beta_{1}}$) and induced ($\mathrm{\beta_{2}}$) solar wind effects likely corresponds with the highly variable transition region between the magnetosheath and the ionosphere of Mars \cite{Holmberg2019}.
Future studies should investigate additional intersections between the influence of intrinsic and induced fields and previously understood transitions in magnetic topology and plasma measurements \cite{Xu2017, Holmberg2019}.

It is worth noting the relatively wide spread of the posterior predictive check \add{and the intrinsic (statistical) scatter ($\sigma$) in the model (see Table 1, footnote)}. 
\add{Our interpretation of the relatively high intrinsic scatter but low intercept ($\beta_{3}, \mathrm{Intercept}$) is that there is a large variation of the data around our general trend but not a variation that can be captured by adding another linear parameter.}
\remove[ver2]{We interpret this to mean that we are missing a physical process that results in an observed scatter, or variation around this trend.}
\add[ver2]{We interpret this to mean that we are missing physical process(es) that results in an observed scatter, or variation around this trend.}
\add{This is not that surprising since this is a simple model that specifically excludes any and all effects from temporal dynamics, mainly current systems that are known to alter Mars' magnetic morphology} \cite{Ramstad2020}.
\add{Given the strength of the trend however, we do not expect that this high scatter affects the conclusions presented in this paper regarding the relative important of the fit coefficients and therefore physical processes.}
\remove{This, in addition to the relatively high intrinsic scatter in the model (see Table 1, footnote), and the large interquartile ranges in Figure 5 are all suggestive that our simplistic model is not capturing another physical variable.}
\remove{This is not that surprising as the overall model is quite simplistic, and we estimate that the leading cause of this variance is a result of temporal variations from current systems (Ramstad et al., 2020).}
\remove{This model however is still accurate in describing the overall trend of our realistic data distributions.}

 \begin{table}[ht]
 \caption{Regression Coefficients from Equation \ref{eqn:model} with 95$\%$ Credible Interval Values$^{a}$ from 60 - 85$\%$ of the Bow Shock Distance from \citeA{Gruesbeck2018}}
 \centering
 \begin{tabular}{l r}
 \hline
  Regression Coefficient & Estimated Value \\
 \hline 
   $\mathrm{\beta_{0, Intrinsic}}$  & $-0.023^{+0.023}_{-0.023}$  \\
   $\mathrm{\beta_{1, Solar Wind}}$ & $ 0.525^{+0.022}_{-0.025}$   \\
   $\mathrm{\beta_{2, Induced}}$    & $ 0.156^{+0.023}_{-0.025}$ \\ 
   $\mathrm{\beta_{3, Intercept}}$  & $ -0.003^{+0.023}_{-0.022}$  \\
   \hline	
   \end{tabular}
   \begin{tablenotes}
   \small{$^{a}$ There is intrinsic scatter in this model with $\sigma = 0.708^{+0.020}_{-0.020}$}. 
 \end{tablenotes}
 \end{table}
 
\note{This table has been updated with additional significant figures, it does not change the interpretation or findings in this paper.}
%----------------------------------------------------------------------------------
\section{Discussion and Conclusion}

In this work, we show that magnetic fields on Mars' dayside exhibit a deformation along $\hat Z$ that corresponds to an induced effect from the IMF direction at high altitudes and to Mars' intrinsic crustal fields at low altitudes. 
This appears as an altitude dependent twist in the draping field over the dayside that propagates downtail with high altitude draping mapping to outer magnetotail regions and lower altitude draping mapping to inner magnetotail regions. 
Dayside draping at low altitudes exhibits a deformation that aligns with the \add{previously reported} observed twist direction of the inner magnetotail \cite{DiBraccio2018, DiBraccio2022} \add{but unlike previous findings which center on reconnection, we find this as a direct result of the crustal field morphology on the dayside.} 
\add{How this dayside morphology from presumably closed crustal fields} \cite<e.g.>[]{Xu2017, Xu2020} \add{is transferred to the tail is an open mystery and the role of reconnection in this transference is unknown.}
\add{This works' results do not include topology which could assist in determining the role of reconnection in transferring the crustal field orientation from the dayside to the tail}
\add[ver2]{ as discussed within} \cite{Xu2020}.
\add{These findings similarly align and explain previously observed dayside draping asymmetries over the northern hemisphere} \cite{Brain2006a, Brain2006b}.
\remove{and of the previously observed dayside draping fields (Brain et al., 2006; Brain, 2006) over the northern hemisphere.}
We \add{additionally} find evidence that the outer magnetotail is twisted as well, but that this is \remove{related} \add{due} to a \add{previously} under quantified induced effect caused by the IMF direction.

\begin{figure}[ht]
 \begin{center}
 \noindent\includegraphics[width=0.9\textwidth]{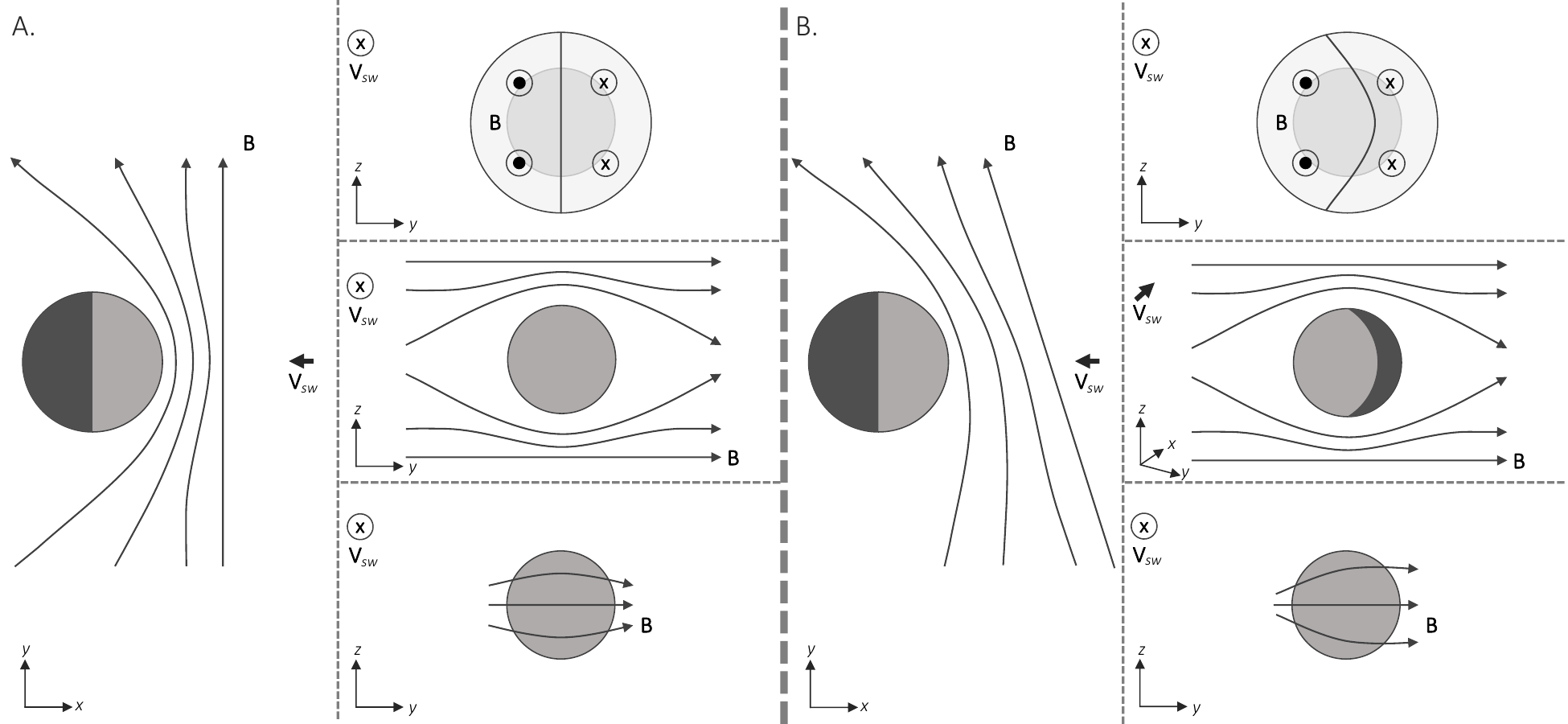}
 \end{center}
 \caption{General induced magnetosphere interaction with expanded IMF directions.
 Panel A shows a case where the solar wind velocity is perpendicular to the IMF. In this case, dayside draping stretches horizontally across the planet and is mostly flat, or parallel to planetary latitude on the dayside. 
 The smaller three figures detail the tail, terminator, and dayside draping regions from upper to lowest figure.
 Panel B shows a case where the IMF additionally has a component in the -X direction, analogous to the $\mathrm {+B_{y}, -B_{x}}$ Parker spiral case at Mars.
 Draping in this situation still stretches horizontally across the planet, but is most parallel to planetary latitude when viewed partially rotated. 
 Instead, when viewed from the dayside, draped field lines will appear twisted, or deflected along $\hat Z$ with the exact deformation depending on the IMF cone angle as a result of draping around a conductive obstacle. 
 These differences in dayside draping by IMF angle results in a symmetric current sheet boundary for panel A, but a crescent shaped current sheet boundary for panel B.
 This schematic does not include deformations \add{or sinking fields} resulting from plume effects as discussed globally in \citeA{Zhang2022} \add{and} \citeA{Chai2019}.}
 \label{fig:schematic}
 \end{figure}

In Figure \ref{fig:schematic} we summarize the impact of this induced effect \add{present in the outer tail and at high dayside altitudes} by showing how the IMF direction alters high altitude magnetic field draping in induced magnetospheres. 
Panels A and B compare an idealized induced magnetosphere for different IMF directions.
Panel A summarizes our traditional understanding of induced magnetospheres where the IMF is perpendicular to the incoming solar wind velocity. 
Panel B summarizes our understanding gathered from this work for when the IMF exhibits a consistent $-\hat X$ component.
This is most analogous to an eastward ($\mathrm{+B_{y}, -B_{x}}$) Parker spiral IMF orientation at Mars. 
In both cases, the incoming solar wind magnetic field must deform around the planetary obstacle as it is conductive.
This can also be seen in the terminator plane in Panel A, and in a plane parallel to the upstream IMF in Panel B as the field in the northern hemisphere deforms northward and the field in the southern hemisphere deforms southward.
This effectively causes what appears to be a twist in the draping fields on the dayside that will deform the outer magnetotail and the current sheet into a crescent shape as the solar wind velocity is still moving downtail.

The Parker spiral angle at Mars creates situations where $\pm \hat X$ directed IMF is common and as a result, we show within this work that Mars' magnetic field draping is often observed to be altered in this way. 
While this induced effect has been shown in physical models \cite{Fang2018}, our understand of the repercussions on dayside draping and global magnetic field morphology have been limited.
\add{For Mars, this general picture we present of induced magnetospheres is limited to high altitudes and the outer tail since the remanent crustal fields dominate the low altitude dayside interaction and inner tail and are not represented in Figure} \ref{fig:schematic}.
\remove{For Mars, this general picture we present of induced magnetospheres is limited at lower altitudes and in the inner tail, since the remanent crustal field dominates the dayside interaction (not represented in Figure 7).}
\add{Similarly, this schematic does not include the effect of field morphology changes as a result of plume dynamics} \cite<e.g. see>{Chai2019, Zhang2022} \add{as it is unlikely that planetary plumes are present in all induced magnetospheres but it is known that this effect is present at Mars}.
\add{The results in this paper estimate that}\remove{this explanation suggests that} distortions along $\hat Z$ in draping fields at high altitudes are a fundamental property of the solar wind interaction, but are altered at lower altitudes and preferentially in the southern hemisphere, due to the intrinsic crustal fields which assists to explain variability in the tail twist \cite<e.g.>{DiBraccio2022}.

These findings point to the draping deformation as an inherent interaction dependent on the incoming solar wind magnetic field geometry, or Parker spiral angle which increases radially from the Sun, and should be variable at other planetary bodies and comets \cite{Parker1958}.
Venus has the most likelihood to demonstrate similar outer tail twisting phenomena to Mars given its distance from the Sun and lack of internal magnetic fields. 
From our findings here, a purely induced effect from the solar wind draping without subsequent alteration from intrinsic fields should result in a crescent shaped current sheet boundary.
Indeed, a crescent shaped boundary was reported at Venus in \citeA{Saunders1986} where they hypothesized that this could have been due to unexplained effects resulting from $\hat X$ directed IMF.
We suggest that this previous observation is the same effect that we quantify and explain within this work as an induced effect of the incoming IMF geometry.

\add[ver3]{Studying magnetotail current sheet responses to the IMF direction is complicated by the potential for both translational (displacement) and rotational (twist) position of the current sheet. At Venus some studies  have been unable to confirm the findings of the displacement of the current sheet discussed within} \citeA{McComas1986} \add[ver3]{even as Mars' tail has shown such displacement} \cite<e.g.>{Rong2016, Romanelli2015}. \add[ver3]{Future work is needed to combine our understanding of current sheet displacement and twist in induced magnetotails.}

Extending these findings to bodies with internal dynamos is complex and how any induced interactions would propagate downtail as discussed within \citeA{Romanelli2022} for Mercury is still unknown. 
Likewise, tail twisting is seen at Earth and has dependence on upstream IMF conditions \cite<e.g.>{Cowley1981, Pitkanen2016} but it is still unclear how this relates to Mars as discussed within \citeA{DiBraccio2022}.
\add{However, recent findings from Earth's magnetosheath show the same morphology dependence on the IMF direction with a far larger data set than is possible for Mars} \cite{MichottedeWelle2022}. 
\citeA{MichottedeWelle2022} \add[ver2]{in particular discusses this effect under the perspective of the repercussions on shock dynamics with changing IMF direction. These repercussions would be worthwhile to integrate with our updated understanding of draping in induced magnetospheres.}\remove[ver2]{in particular discusses this effect under the consideration of shock structures and this viewpoint would be worthwhile to merge with our understanding of induced magnetospheres.}
\add{In general these findings from Venus and Earth further support that the high altitude dayside draping morphology and outer tail twist of Mars as a unifying cross planetary effect.}
We propose that the answer to many of these cross-planetary mysteries of tail twisting and dayside draping then likely lies within first understanding the full three dimensional IMF geometry and its resultant draping configuration with conductive obstacles.
As shown within this work, to first order, this induced solar wind interaction explains the outer tail and high altitude draping while the intrinsic field of Mars explains inner tail and low altitude draping.
It is worthwhile to extend this type of analysis to additional comparative studies to determine how these intrinsic and induced effects exists at other planetary bodies and comets as a fundamental aspect of the solar wind interaction. 

%----------------------------------------------------------------------------------
\section{Open Research}

The MAVEN magnetic field data used in this study are available through the Planetary Data System online at https://pds-ppi.igpp.ucla.edu/mission/MAVEN \cite{PDSMag}. These data are also available at the MAVEN Science Data Center which can be accessed online at https://lasp.colorado.edu/maven/sdc/public/pages/datasets/mag.html. Results generated in this paper benefited from use of the Plotly, pyMC, scikit-learn, and ArviZ software \cite{plotly, arviz, Salvatier2016, scikit-learn}. Data management and workflows for this project benefited from the SQLite (sqlite.org) and the 2i2c projects (2i2c.org).

%%%%%%%%%%%%%%%%%%%%%%%%%%%%%%%%%%%%%%%%%%%%%%%

\acknowledgments
This work was supported by the National Aeronautics and Space Administration (NASA) grant NNH10CC04C to the University of Colorado and by subcontract to Space Sciences Laboratory, University of California, Berkeley. The MAVEN project is supported by NASA through the Mars Exploration Program. We also acknowledge support from NASA's AI / ML Use Case Program, grant 80NSSC21K1370. This work was supported by the NSF Earth Cube Program under awards 1928406, 1928374. E. A. was supported by the National Science Foundation Graduate Research Fellowship under Grant No. DGE 1752814 and the Two Sigma PhD Fellowship. We would like to thank J. Halekas, Y. Dong, B. Langlais, and S. Ruhunusiri, for use of their datasets for solar wind proxies and crustal field models and Y. Panda and E. Sundell of the 2i2c project.

%% ------------------------------------------------------------------------ %%
%% References and Citations

\bibliography{refs}

\end{document}